\documentclass[preprint,aps,showpacs,amsmath,amssymb]{revtex4}
\usepackage{graphicx}
\usepackage{dcolumn}
\usepackage{bm}
\usepackage{mathrsfs}

\renewcommand{\L}{\mathscr{L}}
\newcommand{\M}{\mathscr{M}}
\begin{document}

\title{Decays of $D_{sj}^*(2317)$ and $D_{sj}(2460)$ Mesons in the Quark Model}

\author{Feng-Lin Wang}
\author{Xiao-Lin Chen}
\author{Da-Hai Lu}
\author{Shi-Lin Zhu}
\author{Wei-Zhen Deng}
\email{dwz@th.phy.pku.edu.cn}
\affiliation{Department of Physics, Peking University, Beijing
100871, China}

\begin{abstract}
  We study the decay widths of the narrow resonances $D_{sj}^*(2317)$
  and $D_{sj}(2460)$ in the chiral quark model, together with the well-known
  $D^\ast$ and $D_s^*$ mesons. All the parameters in our calculation are
  taken from Godfrey and Isgur's quark model except the $\pi^0-\eta$
  mixing angle which is fixed by the $D_s^*$ decay widths.  The calculated
  electromagnetic decay widths agree with those from other groups and
  the experimental data available quite well.
  However, the pionic decay widths of $D_{sj}(2317)$ and $D_{sj}(2460)$
  are too small to fit the experimental data. We suspect that the
  simple chiral quark pion axial-vector interaction Hamiltonian is
  not suitable for hadron strong decays of $D_{sj}(2317)$ and $D_{sj}(2460)$.

\end{abstract}

\maketitle

\section{Introduction}

The discovery of narrow resonances $D_{sj}^*(2317)$
\cite{Aubert:2003fg} and $D_{sj}(2460)$ \cite{Besson:2003cp}
raises challenges to the quark model. Masses of these two states
are about 100$ \mbox{MeV}$ lower than the predictions of the
potential quark model \cite{Godfrey:1985xj}. Furthermore, the
isospin conserving decay channels $D_{sj}^{(*)}\to D^{(*)}K$ are
forbidden by kinematics.  The observed pionic decays into
$D_s^{(*)}\pi^0$ break the isospin symmetry.

Several non-conventional schemes, such as molecules
\cite{Barnes:2003dj}, tetraquark states
\cite{Terasaki:2003qa,Cheng:2003kg,Chen:2004dy,Dmitrasinovic:2005gc},
or the chiral partners of $D_s$ and
$D_s^*$ \cite{Bardeen:2003kt,Liu:2006jx} etc., have been proposed (for
a review, see Refs.~\onlinecite{Colangelo:2004vu,Swanson:2006st}).
But the conventional $c\bar{s}$ interpretation is still attractive if
the experimental masses of $D^{(*)}_{sj}$ can be accommodated in the
quark model \cite{Dai:2003yg,Wei:2005ag,Guo:2006fu}. In the heavy
quark limit, $D_{sj}(2317)$ and $D_{sj}(2460)$ naturally form a
$P$-wave doublet $J^P=(0^+,1^+)$ with $j_l = \frac12$. Couple-channel
effects could lead to mass shifts
\cite{vanBeveren:2003kd,Simonov:2004ar}. The observed pionic decays
can also be understood through $\eta-\pi^0$ mixing from the chiral
perturbation theory \cite{Gasser:1984gg,Cho:1994zu}.

In a previous work \cite{Lu:2006ry}, we have calculated the pionic
decay widths of the $D_{sj}^{(*)}$ mesons using the ${}^3P_0$
strong decay model. Another simple decay model to deal pionic
decays is the chiral quark model \cite{Lahde:2002fe}.

In this work, we calculate both the electromagnetic (EM) decays
and strong decay widths of $D_{sj}^{(*)}$ mesons in the quark
model. The meson wave functions are taken from the well-known
Godfrey and Isgur's model \cite{Godfrey:1985xj} which gives an
impressingly good overall description of meson states. The chiral
couplings of light quarks with $\pi$, $\eta$ meson and the isospin
violating $\pi^0$-$\eta$ mixing parameter are taken from the
chiral perturbation theory \cite{Weinberg:1991gf,Cho:1994zu}.
Since the relative momentum is very large in the decays of
$D_{sj}^{(*)}$ mesons, the relativistic effect should be
important. We do not make the non-relativistic reductions of the
transition operators. Instead we keep Dirac spinors in our
calculation.

In the next section, we present our formalism. The decays of
$D_{sj}^{(*)}$ are evaluated and compared with experimental data
in Sec.~III. We give a brief discussion and summary in Sec.~IV.

\section{Decays of $D_{sj}^{(*)}$ in the Quark Model}

The electromagnetic interaction is standard
\begin{equation}
\L_\gamma = e \bar\Psi \gamma^\mu \Psi A_\mu\,.
\end{equation}
In the chiral quark model, the pions interact with the light up
and down quarks through the axial-vector coupling
\begin{equation}
  \L_{\pi q} = \frac{g_A^q}{2f_\pi} \bar\Psi \gamma_\mu \gamma_5 \vec{\tau} \Psi
  \cdot \partial^\mu \vec{\phi}_\pi\,.
\end{equation}
As pointed out in Refs~\onlinecite{Weinberg:1991gf,Cho:1994zu},
the reasonable value of $g_A^q$ ranges from $0.75$ to $1$.

The pionic decay of $D_{sj}^{(*)}$ cannot occur from the above
isospin conserving interaction since $D_{sj}^{(*)}$ has no $u-$ or
$d-$ light flavor quarks in the conventional $c\bar{s}$
configuration. However, this isospin breaking decay can occur
through $\pi^0$-$\eta$ mixing due to the up and down quark mass
difference. In the chiral perturbation theory, the $\eta-\pi^0$
mixing amplitude reads \cite{Gasser:1984gg,Cho:1994zu}
\begin{equation}
\theta_m = \frac{\sqrt3}{4} \frac{m_d - m_u}{m_s -
\frac{m_d + m_u}{2}} \simeq 0.010\,.
\end{equation}
From $SU_F(3)$ symmetry, the $\eta$ coupling should be
\begin{equation}
\L_{\eta q} = \frac{g_A^q}{2 f_\eta} \bar\Psi \gamma_\mu \lambda_8 \Psi
\cdot \partial^\mu \phi_\eta\,.
\end{equation}
Thus, we have \footnote{Here the mixing angle between $\lambda_8$
and $\lambda_0$ can be neglected since it is too small in our
  discussion of isospin violating decays.}
\begin{equation}
\L_{\pi s} = - \frac{g_A^q}{\sqrt3 f_\eta} \theta_m \bar{s} \gamma^\mu \gamma_5
s \partial_\mu \pi^0\,.
\end{equation}

In our calculation, we do not make the non-relativistic reductions of
the transition operators. The quark fields are kept in the form with
Dirac spinors
\begin{equation}
\Psi(x) = \sum_s \int \frac{d^3p}{(2\pi)^3} \frac1{2p^0}
\left[ d^\dag(\bm{p},s) v(\bm{p},s) e^{ip\cdot x}
+ b(\bm{p},s) u(\bm{p},s) e^{-ip\cdot x} \right]\,,
\end{equation}
where the anti-commutation relation of quarks reads
\begin{equation}
\{ d(\bm{p},s), d^\dag(\bm{p}',s') \} = \{ b(\bm{p},s),
b^\dag(\bm{p}',s') \} = (2\pi)^3 (2p^0) \delta_{ss'}
\delta(\bm{p} - \bm{p}')\,.
\end{equation}
Accordingly, the meson wave functions are expressed with the quark
operators
\begin{equation}
|M, \bm{P}\rangle = \frac1{\sqrt{N_c}}
\int \frac{d^3k}{(2\pi)^3} \frac1{2\sqrt{p_Q^0 p_{\bar q}^0}}
\phi(\bm{k},\bm{P},D_s) b^\dag_{Q}
T_M d^\dag_{\bar q} |0\rangle\,.
\end{equation}
Here we treat the quark operators in a matrix form
\begin{align}
b^\dag_Q &= \binom{b^\dag_{Q,\uparrow}}{b^\dag_{Q,\downarrow}}^T
\,, & d^\dag_{\bar q} &= \binom{b^\dag_{\bar
q,\uparrow}}{b^\dag_{\bar q,\downarrow}}\,.
\end{align}
The inner $Q\bar{q}$ structure matrix $T_M$ of $D_s$, $D_s^*$,
$D_{sj}^*(2317)$ and $D_{sj}(2460)$ mesons are taken from the
quark model \cite{Godfrey:1985xj,Godfrey:2005ww}. They are listed
in Table \ref{tab-1}.
\begin{table}
\caption{\label{tab-1}%
The inner tensor structure matrix of $D_s^{(*)}$ mesons.}
\begin{ruledtabular}
\begin{tabular}{lll}
Meson & $J^P$ & $T_M$ \\\hline
$D_s$ & $0^-$ & $\dfrac{\openone}{\sqrt2}$ \\
$D_s^*$ & $1^-$ & $\dfrac{\bm{\sigma}\cdot\bm{\epsilon}}{\sqrt2}$ \\
$D_{sj}^*(2317)$ & $0^+$ & $\dfrac{\bm{\sigma}\cdot\bm{k}}{\sqrt6}$ \\
$D_{sj}(2460)$ & $1^+$ & $\dfrac{\bm{\sigma}\cdot\bm{\epsilon}
\bm{\sigma}\cdot\bm{k}}{\sqrt6}$ \\
${}^1P_1$ & $1^+$ & $\dfrac{\bm{k}\cdot \bm{\epsilon}}{\sqrt2}$ \\
${}^3P_1$ & $1^+$ & $\dfrac{i\bm{k}\times \bm{\sigma} \cdot
\bm{\epsilon}}{2}$
\end{tabular}
\end{ruledtabular}
\end{table}
The spatial wave functions $\phi(\bm{k},\bm{P})$ are normalized as
\begin{equation}
2E = \int \frac{d^3k}{(2\pi)^3} |\phi(\bm{k},\bm{p})|^2\,.
\end{equation}
In the heavy quark limit, $j_l=L+s_q$ is a good quantum number
when $m_Q \to \infty$. The lowest $0^+$ and $1^+$ excitation
states $D_{sj}^*(2317)$ and $D_{sj}(2460)$ have $j_l=\frac12$.
They form the $(0^+, 1^+)$ doublet of the $P$-wave orbital
excitation. $D_{sj}(2460)$ is an ideal mixture of the ${}^1P_1$
and ${}^3P_1$ states in this limit
\begin{equation}
|D_{sj}(2460), \bm{P}\rangle =
\sqrt{\frac13}|D_{sj}({}^1P_1), \bm{P}\rangle +
\sqrt{\frac23}|D_{sj}({}^3P_1), \bm{P}\rangle\,.
\end{equation}

The spatial wave functions are related to the simple harmonic
oscillator (SHO) wave functions in
Ref.~\onlinecite{Godfrey:1985xj}
\begin{equation}
\label{wvf_os}
\frac1{\sqrt{2E}} \phi(\bm{k},\bm{P}) = \phi(\bm{k}^2)\,.
\end{equation}
For the ground states,
\begin{equation}
\phi(\bm{k}^2) = \left( \frac{2\sqrt{\pi}}{\beta} \right)^{3/2}
e^{-\bm{k}^2/2\beta^2}\,.
\end{equation}
For the $P$-wave states,
\begin{equation}
\phi(\bm{k}^2) = \frac{\sqrt2}{\beta}
\left( \frac{2\sqrt{\pi}}{\beta} \right)^{3/2}
e^{-\bm{k}^2/2\beta^2}\,.
\end{equation}

In the rest frame of $A$, the decay width of a process $A\to B+C$ is
\begin{align}
\Gamma(A\to B+C) &= \frac{|\bm{p}_C|}{32\pi^2 M_A^2}
\int d\Omega \left| \M(A\to B+C) \right|^2\,, \\
\M(A\to B+C) &= \langle BC| \L(0) |A\rangle\,,
\end{align}
with
\begin{equation}
\left| \bm{p}_C \right| =
\frac{\left[ (M_A^2 - (M_B + M_C)^2)(M_A^2-(M_B-M_C)^2) \right]^{1/2}}{2M_A}\,.
\end{equation}
The wave function in Eq.~(\ref{wvf_os}) is calculated in the rest
frame, which is an approximation valid only for small $|\bm{P}|$ for
mesons in motion. The calculation of the Lorentz invariant $\M$ matrix
element should be done in a suitable frame in which the relativistic
effect due to $|\bm{P}|$ is small. In our calculation, the $C$
particle is $\pi$ or $\gamma$, which is treated as an elementary
particle. We calculate the invariant matrix element $\M$ in the
frame $\bm{P}_A + \bm{P}_B = 0$ very like the Breit frame. If the heavy
quark is the spectator, the relevant kinematics are
\begin{align}
\bm{P}_A &= \frac12 \bm{p}_C' \\
\bm{P}_B &= -\frac12 \bm{p}_C' \\
\bm{p}_{\bar q,A} &= \bm{k} + \frac12 \bm{p}'_C \\
\bm{p}_{\bar q,B} &= \bm{k} - \frac12 \bm{p}'_C \\
\bm{p}_{Q,A} &= -\bm{k}\\
\bm{p}_{Q,B} &= -\bm{k}\\
\bm{k}_A &= \frac{m_Q\bm{p}_{\bar q,A} - m_q\bm{p}_{Q,A}}{m_q+m_Q}
=\bm{k}+\frac12 \eta_Q \bm{p}'_C \\
\bm{k}_B &= \frac{m_Q\bm{p}_{\bar q,B} - m_q\bm{p}_{Q,B}}{m_q+m_Q}
=\bm{k}-\frac12 \eta_Q \bm{p}'_C
\end{align}
where
\begin{align}
\eta_Q &= \frac{m_Q}{m_Q + m_q}\,, \\
\eta_q &= 1-\eta_Q\,,
\end{align}
and the momentum of $C$ particle in this frame is
\begin{equation}
  \bm{p}'_C = \frac{ (M_A^2 - (M_B + M_C)^2)(M_A^2-(M_B-M_C)^2) }
  {2M_A^2+2M_B^2-M_C^2}\,.
\end{equation}

The matrix elements are listed below. Following
Ref.~\onlinecite{Henriksson:2000gk}, the formulae are all written
in the way similar to the non-relativistic formulae except the
overlapping integrals which approach unity in the non-relativistic
limit $m \to \infty$ and deviate from unity significantly for the
light quarks.

For the pionic decay of the $D^* $ meson, we have
\begin{equation}
  \frac{\langle P\pi |\L(0)| V\rangle}{\sqrt{4E_VE_P}}
  = i \frac{g_A^q}{2f_\pi} F_1(\bm{p}_\pi^2,m_q,\eta_Q)
  \bm{p}_\pi \cdot \epsilon_V\,.
\end{equation}
Its radiative decay amplitude reads
\begin{equation}
\label{VtoP_gamma}
\frac{\langle P\pi| \L_\gamma(0)|V\rangle}{\sqrt{4E_VE_P}} =
i \bm{\epsilon}_V \cdot \bm{p}_\gamma \times \bm{\epsilon}_\gamma
\left[\mu_{\bar q} F_3(\bm{p}_\gamma^2,m_q,\eta_Q)
-\mu_Q F_3(\bm{p}_\gamma^2,m_Q,\eta_q) \right]\,.
\end{equation}
In the case of $D_s^*$, the following substitution is understood
\begin{equation}
 \frac{g_A^q}{2f_\pi} \to - \theta_m \frac{g_A^q}{\sqrt3f_\eta}\,.
\end{equation}

$D_{sj}^*(2317)$ can decay into $D_s$ through the emission of one
$\pi$. The decay matrix element contains two terms
\begin{align}
\frac{\langle P\pi |\L(0)| S\rangle}{\sqrt{4E_SE_P}}
=& -i \sqrt{\frac32} \frac{g_A}{2f_\pi}
\frac{E_\pi \beta}{m_q} F_2(\bm{p}_\pi^2,m_q,\eta_Q)
\notag \\
&+i \eta_Q\sqrt{\frac16} \frac{g_A}{2f_\pi}
\frac{\bm{p}_\pi^2}{\beta} F_1(\bm{p}_\pi^2,m_q,\eta_Q)\,.
\end{align}
Its radiative decay matrix elements also contain two pieces
\begin{align}
  \langle V\gamma |\L_\gamma(0)|S \rangle
&= \langle V\gamma
  |\L_\gamma(0)|S \rangle_E +
  \langle V\gamma |\L_\gamma(0)|S \rangle_M \; ,\\
  \frac{\langle V\gamma |\L_\gamma(0)|S \rangle_E}{\sqrt{4E_VE_S}}
&= \sqrt{\frac23} \beta
  \bm{\epsilon}_V^* \cdot \bm{\epsilon}_\gamma \left[\mu_{\bar q}
    F_4(\bm{p}_\gamma^2,m_q,\eta_Q) -\mu_Q
    F_4(\bm{p}_\gamma^2,m_Q,\eta_q) \right]\; ,\\
  \frac{\langle V\gamma |\L_\gamma(0)|S \rangle_M}{\sqrt{4E_VE_S}}
&= \sqrt{\frac16}
  \frac{\bm{p}^2_\gamma}{\beta} \bm{\epsilon}_V^* \cdot
  \bm{\epsilon}_\gamma \left[\mu_{\bar q}
    \eta_Q F_3(\bm{p}_\gamma^2,m_q,\eta_Q) -\mu_Q
    \eta_q F_3(\bm{p}_\gamma^2,m_Q,\eta_q) \right]\,.
\end{align}

The decay matrix elements of $D_{sj}(2460)$ are complicated, which
are listed according to its decay modes below.
\begin{itemize}
\item $A \to V+\pi$
\begin{align}
\frac{\langle V\pi |\L(0)| A\rangle}{\sqrt{4E_VE_A}}
=& -i \sqrt{\frac32} \frac{g_A}{2f_\pi}
\frac{E_\pi \beta}{m_q} F_2(\bm{p}_\pi^2,m_q,\eta_Q)
\bm{\epsilon}_V^*\cdot\bm{\epsilon}_A
\notag \\
&+i \eta_Q\sqrt{\frac16} \frac{g_A}{2f_\pi}
\frac{\bm{p}_\pi^2}{\beta} F_1(\bm{p}_\pi^2,m_q,\eta_Q)
\bm{\epsilon}_V^*\cdot\bm{\epsilon}_A \,.
\end{align}
\item $A\to P+\gamma$
\begin{align}
  \langle P\gamma |\L_\gamma(0)|A \rangle &= \langle P\gamma
  |\L_\gamma(0)|A \rangle_E +
  \langle P\gamma |\L_\gamma(0)|A \rangle_M\,, \\
  \frac{\langle P\gamma |\L_\gamma(0)|A \rangle_E}{\sqrt{4E_PE_A}}
 &= \sqrt{\frac23} \beta
  \bm{\epsilon}_A \cdot \bm{\epsilon}_\gamma \left[\mu_{\bar q}
    F_4(\bm{p}_\gamma^2,m_q,\eta_Q) -\mu_Q
    F_4(\bm{p}_\gamma^2,m_Q,\eta_q) \right]\\
  \frac{\langle P\gamma |\L_\gamma(0)|A \rangle_M}{\sqrt{4E_PE_A}}
&= \sqrt{\frac16}
  \frac{\bm{p}^2_\gamma}{\beta} \bm{\epsilon}_A \cdot
  \bm{\epsilon}_\gamma \left[\mu_{\bar q}
    \eta_Q F_3(\bm{p}_\gamma^2,m_q,\eta_Q) +\mu_Q
    \eta_q F_3(\bm{p}_\gamma^2,m_Q,\eta_q) \right] \,.
\end{align}
\item $A\to V+\gamma$
\begin{align}
  \langle V\gamma |\L_\gamma(0)|A \rangle &= \langle V\gamma
  |\L_\gamma(0)|A \rangle_E +
  \langle V\gamma |\L_\gamma(0)|A \rangle_M\,, \\
  \frac{\langle V\gamma |\L_\gamma(0)|A \rangle_E}{\sqrt{4E_VE_A}} &=
  i\sqrt{\frac23} \beta \bm{\epsilon}_V^* \times \bm{\epsilon}_A \cdot
  \bm{\epsilon}_\gamma \left[\mu_{\bar q}
    F_4(\bm{p}_\gamma^2,m_q,\eta_Q) -\mu_Q
    F_4(\bm{p}_\gamma^2,m_Q,\eta_q) \right]\\
  \frac{\langle V\gamma |\L_\gamma(0)|A \rangle_M}{\sqrt{4E_VE_A}} &=
  i\eta_Q\sqrt{\frac16} \frac{\bm{p}^2_\gamma}{\beta} \mu_{\bar q}
  \bm{\epsilon}_V^* \times \bm{\epsilon}_A \cdot \bm{\epsilon}_\gamma
  F_3(\bm{p}_\gamma^2,m_q,\eta_Q)  \notag\\
  &-i\eta_q \sqrt{\frac23} \frac{\mu_Q}{\beta}
  F_3(\bm{p}_\gamma^2,m_Q,\eta_q) \bm{\epsilon}_V^* \cdot \left(
    \bm{p}_\gamma \bm{p}_\gamma -\frac12 \bm{p}_\gamma^2 \openone
  \right) \cdot \bm{\epsilon}_A \times \bm{\epsilon}_\gamma\,.
\end{align}
\item $A\to S+\gamma$
\begin{align}
\langle S\gamma |\L_\gamma(0)|A \rangle &=
\langle S\gamma |\L_\gamma(0)|A \rangle_E +
\langle S\gamma |\L_\gamma(0)|A \rangle_M\,, \\
\frac{\langle S\gamma |\L_\gamma(0)|A \rangle_E}{\sqrt{4E_AE_S}} &=
i\left[\frac23(1+\eta_Q) \mu_{\bar q}F_4(\bm{p}_\gamma^2,m_q,\eta_Q)
- \mu_{\bar q}F_5(\bm{p}_\gamma^2,m_q,\eta_Q) \right. \notag \\
&\left. + \frac23 \eta_q \mu_Q F_4(\bm{p}_\gamma^2,m_Q,\eta_q)
- \mu_Q F_5(\bm{p}_\gamma^2,m_Q,\eta_q) \right]
\bm{p}_\gamma \times \bm{\epsilon}_\gamma \cdot \bm{\epsilon}_A \\
\frac{\langle S\gamma |\L_\gamma(0)|A \rangle_M}{\sqrt{4E_AE_S}} &=
 i \frac{\bm{p}_\gamma^2}{6\beta^2} [\eta_Q^2 \mu_{\bar q}
F_3(\bm{p}_\gamma^2,m_q,\eta_Q)+\eta_q^2 \mu_Q F_3(\bm{p}_\gamma^2,m_Q,\eta_q)]
\bm{p}_\gamma \times \bm{\epsilon}_\gamma \cdot \bm{\epsilon}_A\,.
\end{align}

\end{itemize}

\section{Numerical Results}

In our calculation, the wave function parameter $\beta$ and the
quark masses are taken from Ref.~\onlinecite{Godfrey:1985xj}:
\begin{align}
&\beta=400\mbox{MeV}\,, &&m_u=m_d=220\mbox{MeV}\,, &&m_s=419
 \mbox{MeV}\,, &&m_c=1628 \mbox{MeV}\,.
\end{align}
The value of $g_A^q=0.87$ is taken from
Ref.~\onlinecite{Henriksson:2000gk}. We present the pionic decay
widths of $D^{(*)}$ mesons in Table~\ref{tab-2}. Our results agree
with the experimental data very well.
\begin{table}
  \caption{\label{tab-2}%
    Decay widths of $D^* \to D+\pi$ in unit of $\mbox{MeV}$
    and the relevant $F_1$ values.}
\begin{ruledtabular}
\begin{tabular}{lddddd}
$V\to P+\pi$ & \multicolumn{1}{c}{$\bm{p}_\pi$} &
\multicolumn{1}{c}{Exp. \cite{Eidelman:2004wy}} &
\multicolumn{1}{c}{$F_1$} & \multicolumn{1}{c}{Present work} &
\multicolumn{1}{c}{Ref.~\onlinecite{Henriksson:2000gk}} \\\hline
$D^{*\pm} \to D^{\pm}+\pi^0$ & 38 & 0.030 & 0.65 & 0.028 & 0.029 \\
$D^{*\pm} \to D^0 + \pi^\pm$ & 40 & 0.065 & 0.65 & 0.061 & 0.064 \\
$D^{*0} \to D^0 + \pi^0$ & 43 & <1.3 & 0.65 & 0.039 & 0.041
\end{tabular}
\end{ruledtabular}
\end{table}

In Table~\ref{tab-3}, we collect the numerical results of the
isospin violating pionic decays of $D_s^*$ and $D_{sj}^{(*)}$
states. We used the commonly accepted value $\theta_m=0.010$ for
$\pi^0$-$\eta$ mixing.
\begin{table}
\caption{\label{tab-3}%
Pionic decay widths of $D_s^*$ and $D_{sj}^{(*)}$ in unit of
$\mbox{keV}$ and relevant $F_i$ values.}
\begin{ruledtabular}
\begin{tabular}{ldddd}
& \multicolumn{1}{c}{$\bm{p}_\pi(\mbox{MeV})$} &
\multicolumn{1}{c}{$F_1$}
 & \multicolumn{1}{c}{$F_2$} & \multicolumn{1}{c}{Present work} \\\hline
$D_s^* \to D_s + \pi^0$ & 49 & 0.80 && 7.4\times 10^{-3} \\
$D_{sj}^*(2317) \to D_s + \pi^0$ & 298 & 0.70 & 0.52 & 1.9 \\
$D_{sj}(2460) \to D_s^* + \pi^0$ & 297 & 0.70 & 0.52 & 1.9
\end{tabular}
\end{ruledtabular}
\end{table}
The radiative decay widths are listed in Table~\ref{tab-4}
together with some results from other groups.
\begin{table}
\caption{\label{tab-4}%
Radiative decay widths in unit of $\mbox{keV}$.}
\begin{ruledtabular}
\begin{tabular}{ldddddrd}
& \multicolumn{1}{c}{Present work} & \multicolumn{1}{c}{QM
\cite{Ivanov:1994ji}} & \multicolumn{1}{c}{QM \cite{Goity:2000dk}}
& \multicolumn{1}{c}{QM \cite{Godfrey:2003kg}} &
\multicolumn{1}{c}{QSR \cite{Zhu:1996qy}} &
\multicolumn{1}{c}{LCSR \cite{Colangelo:2005hv}} &
\multicolumn{1}{c}{VMD \cite{Colangelo:2004vu}} \\\hline
$D^{*\pm} \to D^\pm\gamma$ & 0.25 & 0.36 & 0.050 & & 0.23 && \\
$D^{*0} \to D^0 \gamma$ & 14.5 & 17.9& 7.3 & & 12.9 && \\
$D_s^* \to D_s \gamma$ & 0.065 & 0.118 & 0.101 & & 0.13 && \\
$D^{*}_{sj}(2317) \to D_s^* \gamma$ &  1.5 & & & 1.9 & & $4-6$ & 0.85 \\
$D_{sj}(2460) \to D_s \gamma$ & 6.3 & & & 6.2 & & $19-29$ & 3.3 \\
$D_{sj}(2460) \to D_s^* \gamma$ & 3.7 & & & 5.5 & & $0.6-1.1$ & 1.5 \\
$D_{sj}(2460) \to D_{sj}^*(2317)$ & 0.18 & & & 0.012 & & $0.5-0.8$ &
\end{tabular}
\end{ruledtabular}
\end{table}
The relevant overlapping integrals are collected in
Table~\ref{tab-6}, where
\begin{align}F_i^q &= F_i(\bm{p}_\gamma^2,m_q,\eta_Q) \\
F_i^Q &= F_i(\bm{p}_\gamma^2,m_Q,\eta_q)
\end{align}
From the table, we see clearly that the overlapping integrals,
which are related to the relativistic effects, are very important
for the light quarks. For the heavy quarks, the overlapping
integrals always approach unity.
\begin{table}
\caption{\label{tab-6}%
Overlapping integrals related to radiative decays.}
\begin{ruledtabular}
\begin{tabular}{ldddddd}
& \multicolumn{1}{c}{$F_3^q$}
& \multicolumn{1}{c}{$F_3^Q$}
& \multicolumn{1}{c}{$F_4^q$}
& \multicolumn{1}{c}{$F_4^Q$}
& \multicolumn{1}{c}{$F_5^q$}
& \multicolumn{1}{c}{$F_5^Q$} \\\hline
$D^{*\pm} \to D^\pm\gamma$ & 0.39 & 0.94 & & & & \\
$D^{*0} \to D^0 \gamma$ & 0.39 & 0.94 & & & & \\
$D_s^* \to D_s \gamma$ & 0.62 & 0.94 & & & & \\
$D^{*}_{sj}(2317) \to D_s^* \gamma$ & 0.60 & 0.94 & 0.56 & 0.93 & & \\
$D_{sj}(2460) \to D_s \gamma$ & 0.47 & 0.92 & 0.44 & 0.91 & & \\
$D_{sj}(2460) \to D_s^* \gamma$ & 0.54 & 0.93 & 0.51 & 0.92 & & \\
$D_{sj}(2460) \to D_{sj}^*(2317)\gamma$ & 0.62 & 0.94 & 0.58 &
0.93 & 0.50 & 0.91
\end{tabular}
\end{ruledtabular}
\end{table}

\section{Summary}

At present, only the decay widths of $D^\ast$ mesons have been
measured. For $D_s^\ast, D_{sj}(2317), D_{sj}(2460)$ states, the
ratios between their radiative and pionic decay widths are
available experimentally, which are collected in
Table~\ref{tab-5}. The experimental data are taken from ``Review
of Particle Physics'' by Particle Data Group (PDG)
 \cite{Eidelman:2004wy} and its online update server:
\texttt{http://pdg.lbl.gov/pdg.html}.

Our calculated ratios $\dfrac{D^{*0} \to D^0 \gamma}{D^{*0} \to
D^0 \pi^0}$ and $\dfrac{D_s^* \to D_s \pi^0}{D_s^* \to D_s
\gamma}$ agree with the experimental ratio within a factor of two.
Such an agreement is quite interesting if we keep in mind both our
model wave function and the strong decay Hamiltonian are so
simple.

Our calculated ratio $\dfrac{D^{*\pm}\to D^\pm\gamma}{D^{*\pm} \to
  D^\pm\pi^0}$ is nearly six times smaller than the experimental data.
This may be partly attributed to the uncertainty of our naive SHO
wave functions. In the $V\to P+\gamma$ formula Eq.
(\ref{VtoP_gamma}), there exists a strong cancellation between the
light and heavy quark contributions. We have
\begin{equation}
\frac{\mu_{\bar d}}{\mu_c} = \frac{m_c}{2 m_d} \approx -
\frac{1600}{440}\,,
\end{equation}
and
\begin{equation}
\frac{F_3^q}{F_3^Q} \approx 0.4\,,
\end{equation}
i.e., $\mu_{\bar d}F_3^q \approx - \mu_c F_3^Q$ which leads to the
strong cancellation. The sensitivity of the overlapping integrals
to the uncertainty of the wave function is amplified in this case. For
example, if we change the $\beta$ parameter to
$\beta=300\mbox{MeV}$, we have $F_3^q \approx 0.48$. Then the
resulting ratio will increase to $\sim 0.02$.

As can be seen in Table~\ref{tab-4}, the radiative decay widths of
different channels in this work are comparable with those from other
groups (see also ref.~\onlinecite{Ebert:2002xz}). However, the pionic
decay widths of $D_{sj}(2317)$ and $D_{sj}(2460)$ from the simple
chiral quark model in Table~\ref{tab-3} are ten times
smaller than those from light-cone QCD sum rules approach
\cite{Wei:2005ag} and the $^3P_0$ decay model \cite{Lu:2006ry}.  Hence
our calculated ratios between EM and pionic decay widths of
$D_{sj}(2317)$ and $D_{sj}(2460)$ mesons are systematically larger
than the experimental data by a factor of $10$.  Such a large
systematic discrepancy cannot easily be ascribed to either the
uncertainty of the meson wave function or the uncertainty of the value
of the $\eta-\pi^0$ mixing amplitude \cite{Coon:1981qs,Lahde:2002fe}.
We tend to conclude that the simple strong decay mechanism based on
the pion and chiral quark axial vector coupling is not realistic if
$D_{sj}(2317)$ and $D_{sj}(2460)$ mesons are conventional $c\bar s$
states.

\begin{table}
\caption{\label{tab-5}%
Branching ratios between radiative and pionic decays.}
\begin{ruledtabular}
\begin{tabular}{ldd}
& \multicolumn{1}{c}{Exp.} & \multicolumn{1}{c}{Present work}
\\\hline
$\dfrac{D^{*0} \to D^0 \gamma}{D^{*0} \to D^0 \pi^0}$ &
0.62 & 0.35 \\
$\dfrac{D_s^* \to D_s \pi^0}{D_s^* \to D_s \gamma}$ &
0.062 & 0.11 \\
$\dfrac{D^{*\pm} \to D^\pm \gamma}{D^{*\pm} \to D^\pm \pi^0}$ &
0.052 & 0.009 \\
$\dfrac{D_{sj}^*(2317) \to D_s^* \gamma}{D_{sj}^*(2317) \to D_s
\pi^0}$ &
<0.059 & 0.79 \\
$\dfrac{D_{sj}(2460) \to D_s \gamma}{D_{sj}(2460) \to D_s^*
\pi^0}$ &
0.31 & 3.3 \\
$\dfrac{D_{sj}(2460) \to D_s^* \gamma}{D_{sj}(2460) \to D_s^*
\pi^0}$ &
<0.16 & 1.9 \\
$\dfrac{D_{sj}(2460) \to D_{sj}^*(2317) \gamma}{D_{sj}(2460) \to
D_s^* \pi^0}$ & <0.22 & 0.094
\end{tabular}
\end{ruledtabular}
\end{table}

In summary, we perform a systematic calculation of
the decay widths of $D_s^*$, $D_{sj}^*(2317)$, and $D_{sj}(2460)$.
The EM radiative decay widths agree with the available
experimental data and other model results reasonably well.
But the isospin violating pionic decay widths of $D_{sj}^*(2317)$ and
$D_{sj}(2460)$ are too small to fit the experimental data.
This disagreement cannot easily be resolved by changing the wave
functions or the $\eta-\pi^0$ mixing amplitude in the chiral quark
model. One may wonder whether other possible theoretical schemes
such as the coupled-channel effects, hybrid meson, molecule state
or tetraquark interpretations of these two resonances may resolve
the above discrepancy. However, there is no clear evidence in
favor of these exotic schemes from BABAR's most recent extensive
measurement \cite{Aubert:2006bk}. Therefore we strongly suspect the
chiral quark pion interaction Hamiltonian may be too simple to
describe strong decays reliably.

\begin{acknowledgments}

This project was supported by the National Natural Science
Foundation of China under Grants 10375003 and 10421503, Ministry
of Education of China, FANEDD, Key Grant Project of Chinese
Ministry of Education (NO 305001) and SRF for ROCS, SEM.

\end{acknowledgments}

\appendix
\section{The overlapping integrals}
All $F_i(\bm{p}^2,m,\eta)$ can be expressed as
\begin{equation}
F_i(\bm{p}^2,m,\eta) =
e^{-\frac14\eta \bm{p}^2/\beta^2}\left(\frac{2\sqrt{\pi}}{\beta} \right)^3
\int \frac{d^3 \bm{k}}{(2\pi)^3}
e^{-\bm{k}^2/\beta^2} \sqrt{\frac{q^0+m}{2q^0}}
\sqrt{\frac{q^{\prime 0} + m}{2 q^{\prime 0}}} G_i
\end{equation}
where
\begin{align}
  G_1 &= 1 + \frac{2(\bm{k}\cdot \hat{\bm{p}})^2 -
    (\bm{k}^2+\frac14\bm{p}^2)}
  {(q^0 + m)(q^{\prime 0} + m)} \\
  G_2 &= \frac{2\bm{k}^2}{3\beta^2} \left [ \frac{m(q^0+q^{\prime 0} +
      2m)}{(q^0+m)(q^{\prime 0} + m)} -
    \frac{m(\hat{\bm{k}}\cdot\bm{p})^2}{(q^0+m)(q^{\prime 0} +
      m)(q^0+q^{\prime 0})}
  \right] \\
  G_3 &= \frac{m(q^0+q^{\prime 0} + 2m)}{(q^0+m)(q^{\prime 0} + m)} -
  \frac{4m(\hat{\bm{p}}\cdot\bm{k})^2}{(q^0+m)(q^{\prime 0} +
    m)(q^0+q^{\prime 0})} \\
G_4 &= \frac{m(q^0+q^{\prime 0} + 2m)}{(q^0+m)(q^{\prime 0} + m)}
\frac{\bm{k}^2 - (\bm{k}\cdot\hat{\bm{p}})^2}{\beta^2} \\
G_5 &= \frac{2\bm{k}^2}{3\beta^2}
\left[\frac{m(q^0+q^{\prime 0} + 2m)}{(q^0+m)(q^{\prime 0} + m)} -
  \frac{4m(\hat{\bm{p}}\cdot\bm{k})^2}{(q^0+m)(q^{\prime 0} +
    m)(q^0+q^{\prime 0})}\right]
\end{align}
and $\bm{q}$ and $\bm{q}'$ are quark's momenta
\begin{align}
\bm{q} &= \bm{k}+\frac12\bm{p} \\
\bm{q}' &= \bm{k}-\frac12\bm{p}
\end{align}


\end{document}